# Using Affective Features from Media Content Metadata for Better Movie Recommendations

John Kalung Leung[1][a], Igor Griva and William G. Kennedy[2][b]

[1] *Computational and Data Sciences Department, Computational Sciences and Informatics, College of Science, George Mason University, 4400 University Drive, Fairfax, Virginia 22030, USA*

[2] *Department of Mathematical Sciences, MS3F2, Exploratory Hall 4114, George Mason University, 4400 University Drive, Fairfax, Virginia 22030, USA*

[3]*Center for Social Complexity, Computational and Data Sciences Department, College of Science, George Mason University, 4400 University Drive, Fairfax, Virginia 22030, USA*

*{jleung2, igriva, wkennedy}@gmu.edu*

Abstract: This paper investigates the causality in the decision making of movie recommendations through the users' affective profiles. We advocate a method of assigning emotional tags to a movie by the auto-detection of the affective features in the movie's overview. We apply a text-based Emotion Detection and Recognition model, which trained by tweets short messages and transfers the learned model to detect movie overviews' implicit affective features. We vectorize the affective movie tags to represent the mood embeddings of the movie. We obtain the user's emotional features by taking the average of all the movies' affective vectors the user has watched. We apply five-distance metrics to rank the Top-N movie recommendations against the user's emotion profile. We found Cosine Similarity distance metrics performed better than other distance metrics measures. We conclude that by replacing the top-N recommendations generated by the Recommender with the reranked recommendations list made by the Cosine Similarity distance metrics, the user will effectively get affective aware top-N recommendations while making the Recommender feels like an Emotion Aware Recommender.

## 1 INTRODUCTION

Emotion affects human experience and influences our daily activities on all levels of the decision-making process. When a user ponders over a list of recommended items such as songs, books, movies, products, or services, his affective state of preferences influences his decision making on which recommended item he chooses to consume. Emotion plays a role in our decision-making process in preference selection (Naqvi et al., 2006). However, the information retrieval (IF) and Recommender Systems (RS) field give little attention to include human emotion as a source of user context (Ho and Tagmouti, 2006). Our goal in this paper is to make affective awareness a component in making movie recommendations for users. The challenge is that no film database or movie dataset in the public domain contains any explicit textual oriented human emotional tag in the metadata. However, the film metadata fields such as plot, overview, storyline, script, watcher reviews, and critics reviews contain excellent subjective data that describes the general mood of a movie. We can apply Machine Learning (ML) techniques to identify and extract affective features implicitly from the film metadata and leverage the film's emotional characteristics when making movie recommendations to users.

No two films are created the same. The moods of a movie act like an affective fingerprint of the film. We envision an affective movie feature represents by a low dimension continuous emotional vector embedding denotes as the movie's emotional vector (mvec). Some film databases, such as the MovieLens, track users' movie-watching history and feedback (Harper and Konstan, 2016). Using the user's movie-watching history, we formulate a low dimension continuous emotion vector embedding denotes as the user emotional vector (uvec). We obtain a user's uvec embedding value by taking the average of all the movies' mvecs the user has watched. Note that uvec may not be unique if two users watched the same set of movies. The difference between mvec and uvec is that mvec of a movie is static, with value unchanged throughout its lifetime.

---

[a] https://orcid.org/0000-0003-0216-1134

[b] https://orcid.org/0000-0001-9238-1215

In contrast, uvec is dynamic, with its value changes as the user watched and rated a movie. The advantage of using the dynamic nature of uvec in the movie recommendation-making process is that we are taking the most updated user's affective preference into consideration of the user's decision-making process. As the user emotional preference change, the movie recommender will adjust the recommendation-making process accordingly. We may be the first party making use of the novelty in leveraging the dynamic nature of uvec over mvec to enhance the movie Recommender recommendation-making process.

Table 1: Affect values, mvec, of movies "The Godfather (1972)" derived from balanced and unbalanced moods.

| Moods | Balanced Moods Dataset | Rank | Unbalanced Moods Dataset | Rank |
|---|---|---|---|---|
| Neutral | 0.0840931 | 6 | 0.04276474 | 6 |
| Joy | 0.059261046 | 7 | 0.16501102 | 3 |
| Sadness | 0.08991193 | 5 | 0.076094896 | 4 |
| Hate | 0.23262443 | 1 | 0.4305178 | 1 |
| Anger | 0.20177138 | 2 | 0.1993026 | 2 |
| Disgust | 0.19720455 | 3 | 0.053966276 | 5 |
| Surprise | 0.13513364 | 4 | 0.03234269 | 7 |

Moreover, we can leverage the range and strength of a film's moods, i.e., mvec, to analyse a film's emotional features. In this study, we track six primary human affective features in emotion: "joy", "sadness", "hate", "anger", "disgust", and "surprise". We added "neutral" as the seventh affective feature for convenience in affective computation. We normalized the affective features when we compute mvec for a film. Thus, all affective features in mvec will add up to one (1). For example, Internet Movie Database (IMDb) is a popular online movie information database that has rated "The Godfather (1972)" as the top movie of all time (IMDb, 2020). Our emotion detector classified the movie's dominant affective class as "hate" and depicted the movie's mvec in Table 1.

## 2 RELATED WORK

Detecting primary human emotion expression in text is a relatively new research area in Natural Language Processing (NLP). A common approach in identifying the general thought, feeling, or sense in writing is to classify the contextual polarity orientation (positive, neutral, and negative) of opinionated text through the polarity Sentimental Analysis (SA) (Wilson et al., 2005), and (Maas et al., 2011). When applying fine-grained Sentiment Analysis (Fink et al., 2011), researchers can identify the intensity level of the polarity as a multi-class single-label classification problem (e.g., very positive, optimistic, neutral, negative, and very negative) (Bhowmick et al., 2009). However, to determine the mental, emotional state or composure (i.e., mood) in subjective text, Emotional Analysis (EA) can better suit to handle the task (Tripathi et al., 2016). The researcher wants to know the writing feeling under examination is one of the following primary human emotions or moods.

The study of basic human emotional expressions started in the era of Aristotle in around 4th century BC (Konstan and Konstan, 2006). However, not until Charles Darwin (1872 – 1998) revisited the investigation of human emotional expression in the 19th century, which propelled the field to its present stage of modern psychology research (Ekman, 2006). Paul Ekman et alia in the 1970s developed a Facial Action Coding System (FACS) to carry out a series of research on facial expressions that have identified the following six primary universal human emotions: happiness, sadness, disgust, fear, surprise, and anger (Ekman, 1999). Ekman later added contempt as the seventh primary human emotion to his list (Ekman et al., 2013). Robert Plutchik invented the Wheel of Emotions, advocated eight primary emotions: anger, anticipation, joy, trust, fear, surprise, sadness, and disgust. Adding to the primary eight emotions are secondary and complementary emotions for 32 emotions depicted on the initial Wheel of Emotions (Plutchik, 2001). More recent research by Glasgow University in 2014 amended that couple pairs of emotions such as fear and surprise elicited similar facial muscles response, so are disgust and anger. The study broke the raw human emotions down to four fundamental emotions: happiness, sadness, fear/surprise, and disgust/anger (Tayib and Jamaludin, 2016).

Like many researchers have based their work on Ekman's six primary human emotions (Canales and Martínez-Barco, 2014), we also focus our emotion detection and recognition (EDR) on Ekman's six primary human emotions. We make use of the WordNet-Affect, a linguistic resource for a lexical representation of affective knowledge in affective computing on human interaction such as attention, emotions, motivation, pleasure, and entertainment (Valitutti et al., 2004). Emotional expression research usually aims to detect and recognize emotion types from human facial expression and vocal intonation (De Silva et al., 1997). However, our EDR study focuses on the mood of text expression instead. Nevertheless, the question remains how much of an emotion we can convey through writing.

## 3 METHODOLOGY

In the absence of any publicly available explicit emotion labeled movie metadata dataset, we build an affective text aware model in two steps. First, through readily available tweets data from the Twitter database, we developed a Tweet Affective Classifier (TAC). TAC can classify any tweet text into an affective vector embedding containing seven basic human emotions in probabilistic values. Next, we feed the movie text metadata, such as overviews, to TAC to classify the movie's mvec affective values.

### 3.1 Data Preparation

One of the challenges in this study is to obtain a large enough movie metadata set with mood labels. No such dataset is readily available. We need to build the required dataset by deriving it from four different sources:
1. For the movie rating datasets, we obtained the datasets from the MovieLens datasets stored in the GroupLens repository (Harper and Konstan, 2016).
2. We scraped The Movie Database (TMDb) (TMDb, 2018) for movie overviews and other metadata.
3. We derived our emotional word sense set as contextual emotional words synonymous with WordNet (Miller, 1995).
4. Finally, we scraped the Twitter database for tweets with keyword tags that matched our contextual emotion word synonymous (Marres and Weltevrede, 2013).

MovieLens contains a "links" file containing cross-reference links between MovieLens' movie id and TMDb's tmdb id. We connect MovieLens and TMDb datasets through the "links" file.

We build a seven text-based emotion predictor for movie overviews from the seven-emotion tweet classifier model. We run the predictor through all the 452,102 overviews scraped from the TMDb database. We joined the affective aware movie overviews with four Movielens datasets: ml-latest-small (a.k.a. ml-latest-small hereafter), ml20m, ml25m and ml-latest (a.k.a. ml27m hereafter), as depicted in the following Table 2. Not all movies listed in the MovieLens datasets has a corresponding movie extracted from TMDb.

Table 2: MovieLens dataset statistics merged with number of overviews from TMDb.

| Dataset | users | ratings | movies | overviews |
|---|---|---|---|---|
| mlsm | 610 | 100836 | 9742 | 9625 |
| ml20m | 138493 | 20000263 | 27278 | 26603 |
| ml25m | 162641 | 25000095 | 62423 | 60494 |
| ml27m | 283228 | 27753444 | 58098 | 56314 |

#### 3.1.1 Extract emotion synonymous from WordNetAffectEmotionLists

WordNet developed an affective knowledge linguistic resource known as WordNet-Affect for lexical representation (Strapparava et al., 2004). The selection and tagging of a subset of synsets convey the emotional meaning of a word in WordNet-Affect. WordNet-Affect emotion lists contain lists of concepts extracted from WordNet-Affect, synsets with six emotions of interest: anger, disgust, hate, joy, sadness, and surprise stored in a compressed file:

"WordNetAffectEmotionLists.tar.gz" (Poria et al., 2012).

We downloaded the ".gz" file and uncompressed it into six emotion text files. Each emotion file contains two columns of information: synsets and the synonymous. Here, the synonymous set of the synset corresponds to an emotion class and store in the corresponding emotion text file. We extract the synonymous column from each emotion text file. We removed duplicate synonymous, sorted the cleansed synonymous, and stored the result in comma-separated values (CSV) format in the corresponding emotion synonymous file. Table 3 contains the statistic of the six emotion synonymous files after performed the data cleansing task.

Table 3: Synonymous statistics of six emotions.

| Mood | Count | Synonymous List |
|---|---|---|
| Joy | 400 | "admirable",…,"zestfulness" |
| Sadness | 202 | "aggrieve",…,"wretched" |
| Hate | 147 | "affright",…,"unsure" |
| Anger | 255 | "abhor",...,"wrothful" |
| Disgust | 53 | "abhorrent",...,"yucky" |
| Surprise | 71 | "admiration",...,"wondrously" |

### 3.1.2 Extract tweets from Twitter database

There are many types of tweets on Twitter, a popular social network, and the microblogging platform. This study only works with the regular tweet, 140 characters, or less short message, which posts on Twitter. Almost every user's tweets are extractable and available to the public. Each tweet is searchable by keyword. We wrote a simple Python script to extract tweets through Twitter's API (Makice, 2009). We treat each synonymous in an emotion corresponding file as a keyword of a tweet. By looping through all the synonymous in Twitter's search-by-keyword API, we extract all the tweets with such keyword and store them in a CSV file. The alternative is to extract tweets and store them in a JSON file, as illustrated in (Makice, 2009). For example, if the emotion synonymous belongs to the anger concept, we will store the retrieved tweet in *anger raw.csv* file. As depicted in Table 3, the anger emotion corresponding file, the anger syn.txt, has 255 synonymous. We will store all tweets retrieved from the corresponding keywords in *anger raw.csv*.

After performing text cleansing steps, it yields the following affective feature records depicted in Table 4. Our affective dataset extracted from Twitter shows an unbalanced dataset. We decide to balance the affective dataset by subsampling each affective attribute dataset size to 15,000. We further split each affective dataset into a training dataset with 80% of the samples (12,000) and 20% of the test dataset (3,000).

Using a brute force method, we scrape the TMDb database for movie metadata and movie, which contains the subjective writing movie description to classify the text's mood. Our effort yields 452,102 records after cleansing the raw data we scraped from TMDb.

Table 4: Moods records gathered from Twitter.

| Mood Class | Size |
|---|---|
| Neutral | 19180 |
| Joy | 138019 |
| Sadness | 60381 |
| Hate | 38651 |
| Anger | 17830 |
| Disgust | 19887 |
| Surprise | 15002 |
| No. of unbalance 7 mood classes | 308878 |
| No. of each balanced mood | 15000 |
| No. of each balanced mood train | 12000 |
| No. of each balanced mood test | 3000 |

## 3.2 Emotion Modeling

Intrigue by the recent publication in Natural Language Processing (NLP) for text classification described by (Sosa, 2017), we develop a text-based EDR model by concatenating Long Short-Term Memory (LSTM) architecture and Conv-1D of Convolutional Neural Network (CNN) architecture.

We follow a similar method used in (Liu, 2020) to build our EDR model. We define our model architecture consists of two halves. The first half is RNN LSTM-CNN Conv-1D architecture, as described in (Sosa, 2017) that text input process by an LSTM architecture before following up data processing by a CNN Conv-1D architecture. In contrast, the second half of the model is to reverse the processing order of architecture, CNN Conv-1D-RNN LSTM. Input first process by a CNN Conv-1D architecture before feeding it to an RNN LSTM architecture. The two halves of the architecture then combine to feed data into a max-pooling layer of a CNN for a pooling operation to select the dominant feature in the filter's regional feature map. Next, data passes into a CNN flattening layer to convert data into a one-dimensional array before feed data to a fully connected CNN layer. The dense layer's output will feed to a set of nodes equal to the number of classes the architecture aims to classify. Each of the output nodes holds the output distribution value of its class. In the final act, a softmax activation function examine and activate the appropriate class node accordingly.

We use bi-directional RNN LSTM and CNN Conv1D architectures to build our model. In the first half of the model, the RNN LSTM-CNN Conv-1D phase, we use two bi-directional LSTM for the RNN LSTM architecture and seven Conv1D CNN architecture. In the second half of the model, the CNN-LSTM phase, we apply seven pairs of Conv1D of the CNN architecture and two bi-directional LSTM for the LSTM architecture. Follow the idea illustrated in (Kim, 2014); we prepared two identical input layers of embedding matrix constructed from a pre-trained GloVe embedding matric similar to (Pennington et al., 2014). We build the two input layers of the embedding matrix with one of the input embedding layers set to "trainable," while the other is not, i.e., "frozen." During the processing of the first half of the model, the RNN LSTM-CNN Conv-1D phase, the "trainable" input layer occupies one of the bi-directional LSTM architectures. In contrast, the "frozen" input layer fills the other.

Table 5: Emotion Detection Recognition on balanced moods tweets.

|  | Precision | Recall | F1-score | Support |
|---|---|---|---|---|
| Neutral | 0.47 | 0.77 | 0.59 | 2992 |
| Joy | 0.63 | 0.53 | 0.58 | 3030 |
| Sadness | 0.64 | 0.44 | 0.52 | 3034 |
| Hate | 0.64 | 0.51 | 0.57 | 2933 |
| Anger | 0.62 | 0.68 | 0.65 | 2984 |
| Disgust | 0.44 | 0.45 | 0.44 | 2987 |
| Surprise | 0.55 | 0.51 | 0.53 | 3040 |
| Accuracy |  |  | 0.56 | 21000 |
| Macro avg | 0.57 | 0.56 | 0.55 | 21000 |
| Weighted avg | 0.57 | 0.56 | 0.55 | 21000 |

Similarly, when processing the second half of the model, the CNN Conv-1D – RNN LSTM phase, a "trainable" input layer, and a "frozen" input layer will occupy each pair of the Conv1D units. We obtain 55.6% accuracy in classifying the emotion class of the tweets' balanced dataset, as depicted in Table 5, and the confusion matrix in figure 1 depicts the performance of the seven-emotion classifier. The classification result is acceptable to serve our purpose since our goal is not to build the best emotion text classifier, but a usable one to classify the emotion class of movie overviews.

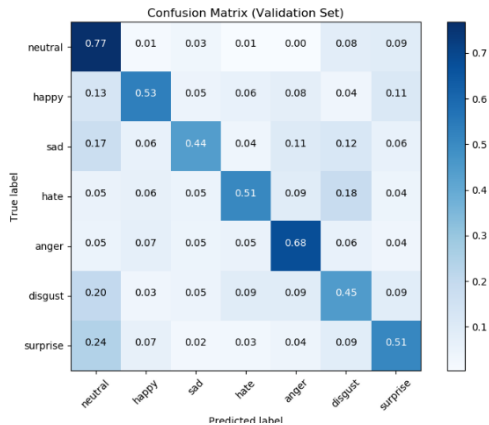

Figure 1: Confusion matrix of 7 balanced emotions dataset.

# 4 IMPLEMENTATIONS

## 4.1 UVEC and MVEC

We selected all users in the MovieLens ml-latest-small dataset as test users. We split the number of movies watched by each user into the user training set and validation set. Table 6 illustrates of the computation results of uvecs from a 20-80% split of train and validation sets. Table 8 depicts an example of user id 400 uvec statistics. We sorted user id 400 by the timestamp attribute in the ml-latest-small ratings data file to simulate the movie watching order. We take the 20% split train set to compute the uvec for the test user. We used the test user's last watched movie, "Lucky Number Slevin (2006)," as input to the SVD-CFRS Recommender to generate a top-20 recommendations list. We applied the five-distance metrics against the top-20 recommendation list to obtain five corresponding re-ranked top-20 recommendation lists, as depicted in Table 8. We then compared the number of movies matched in each recommendation list against the movies in the validation set of the to-be-watched list to report as the hit percentage.

Table 6: MLSM dataset test user uvec, movie watched count, 20% train and 80% validate.

| User ID | 1 | 2 | … | 610 |
|---|---|---|---|---|
| Wcount | 2698 | 1864 | … | 2108 |
| x%train |  |  | … |  |
| y%validate |  |  | … |  |
| UVEC |  |  | … |  |
| Neutral | 0.16635188 | 0.17283309 |  | 0.16885831 |
| Joy | 0.09730581 | 0.09685813 |  | 0.09974659 |
| Sadness | 0.1180924 | 0.11604573 |  | 0.1187206 |
| Hate | 0.1641951 | 0.16120733 |  | 0.16087716 |
| Anger | 0.11517799 | 0.11227607 |  | 0.11261272 |
| Disgust | 0.17250315 | 0.17098578 |  | 0.17191968 |
| Surprise | 0.16637367 | 0.16979389 |  | 0.16726495 |

## 4.2 Comparative Platform

We developed a movie Recommender System to evaluate the performance of user emotion profile, uvec, and movie emotion profile, mvec. We envision uvec and mvec play a role in the tail end process of making movie top-N Recommendations. Any movie Recommenders can refit to support uvec and mvec processes. We developed an SVD based Collaborative Filtering movie Recommender System (SVD-CFRS). We added functions to support uvec and mvec operations as enhancement of making movie recommendations.

The five-distance metrics we employed in the comparative platform were Euclidean distance, Manhattan distance, Minkowski distance, Cosine similarity, and Pearson correlation with their formula illustrated in equation 1 through 6.

$$Euclide(x, y) = \sqrt{\sum_{i=1}^{n}(x_i - y_i)^2} \qquad (1)$$

$$Manhattan(x, y) = \sum_{i=1}^{n} |x_i - y_i| \quad (2)$$

$$Minkowski(x, y) = \left(\sum_{i=1}^{n} |x_i - y_i|^p\right)^{\frac{1}{p}} \quad (3)$$

$$Inner(x, y) = \sum_{i} x_i y_i = \langle x, y \rangle \quad (4)$$

$$CosSim(x, y) = \frac{\sum_i x_i y_i}{\sqrt{\sum_i x_i^2} \sqrt{\sum_i y_i^2}} = \frac{<x, y>}{\|x\| \|y\|} \quad (5)$$

$$PearCorr(x, y) = CosSim(x - \overline{x}, y - \overline{y}) \quad (6)$$

Where (x, y) are vectors $x = (x_1, \ldots, x_n)$ and $y = (y_1, \ldots, y_n)$.

## 5 EVALUATION

### 5.1 Findings

We derived the following methods to evaluate the performance of the recommendations list generated by the SVD-CF Recommender and the reranking of the recommendations list by the five-distance metrics algorithms. We join the ml-latest-small dataset's ratings and movie data files with the emotion TMDb dataset to obtain the movie dataset with emotion labels. Table 7 illustrates the pseudo-code of grouping users and sorted their activity by timestamp to simulate their movie watching order. The pseudo-code also shows splitting each user's activity into a test and validation set and compute uvec. It also shows in the pseudo-code the top-20, top-10, and top-5 of recommendations lists and the reranking of the recommendations list by the five-distance metrics. Lastly, the pseudo-code computes the hit rate in percent for top-20, top-10, and top-5 recommendations.

Table 7: Pseudocode of getting users uvecs, topN, and hit rates.

| 1 | Group user by userId |
|---|---|
| 2 | For each user do |
| 3 |    Sort user row by timestamp |
| 4 |    Split user into test and validation set: 20/80 |
| 5 |    Get uvec for test user |
| 6 |    Get top20/10/5 by test user's last watched film |
| 7 |    Rerank top20 by 5 distance metrics, get top10/5 |
| 8 |    Get hit% by validation set for top20/10/5 |
| 9 |    Get hit% by validation set for 5-dist top20/10/5 |

Table 8 shows an example of test user id 400 of last watched movie statistics and the movie's mvec.

Table 8: Test user id 400 last watched movie and mvec.

| uwerId | 400 | tmdbId | 186 |
|---|---|---|---|
| movieId | 44665 | rating | 4.0 |
| title | Lucky Number Slevin (2006) | genres | Crime, Drama, Mystery |
| timestamp | 1498870148 | m_neutral | 0.204775 |
| m_happy | 0.075458 | m_sad | 0.127180 |
| m_hate | 0.171851 | m_anger | 0.112534 |
| m_disgust | 0.166984 | m_surprise | 0.141217 |

Table 9 depicts a 20-80 split sampling result of the test user id 400. The column "Mid" shows the movie ids of top-20, top-10, and top-5 recommendations list generated by the SVD-CF Recommender and the recommendations lists' reranking by the five-distance metrics. Also, the table shows the hit rates of top-20, top-10, and top-5 for each category.

Table 9: Sampling on 20-80 split of 43 watched movies of ml-latest-small user id 400 into 8 watched movies for uvec computation and 35 to-be-watched movies for validation on the top20 and top5 recommendations list generated by SVD-CFRS and five distance.

| 400 | Mid | Euc | Mht | Mki | Cos | Pear |
|---|---|---|---|---|---|---|
| 1 | 296 | 58559 | 58559 | 58559 | 2959 | 2959 |
| 2 | 50 | 7153 | 7153 | 7153 | 2329 | 58559 |
| 3 | 858 | 356 | 5952 | 356 | 527 | 2028 |
| 4 | 2959 | 5952 | 356 | 5952 | 2858 | 2329 |
| 5 | 593 | 593 | 1089 | 593 | 1213 | 527 |
| 6 | 4993 | 1089 | 593 | 1089 | 858 | 858 |
| 7 | 58559 | 1221 | 1221 | 608 | 79132 | 1089 |
| 8 | 7153 | 608 | 50 | 1221 | 2028 | 2858 |
| 9 | 608 | 4993 | 4993 | 4993 | 47 | 1213 |
| 10 | 1221 | 2028 | 2028 | 2028 | 50 | 47 |
| 11 | 527 | 296 | 608 | 296 | 296 | 79132 |
| 12 | 79132 | 50 | 296 | 47 | 4993 | 4993 |
| 13 | 1213 | 47 | 47 | 79132 | 608 | 1221 |
| 14 | 5952 | 858 | 858 | 50 | 1089 | 7153 |
| 15 | 2858 | 79132 | 79132 | 858 | 1221 | 296 |
| 16 | 356 | 1213 | 2858 | 1213 | 593 | 608 |
| 17 | 47 | 2858 | 527 | 2858 | 5952 | 593 |

| 18 | 2329 | 527 | 2329 | 527 | 356 | 5952 |
| 19 | 1089 | 2329 | 1213 | 2329 | 58559 | 50 |
| 20 | 2028 | 2959 | 2959 | 2959 | 7153 | 356 |
| T20% | 70 | 70 | 70 | 70 | 70 | 70 |
| T5% | 100 | 80 | 60 | 80 | 40 | 40 |
| T10% | 100 | 70 | 70 | 70 | 60 | 50 |

## 5.2 Find the winner of TopN ranked by distance metrics

Table 10 shows the average categorical hit rate of all test users. The first column, "Mid" shows the average hit rate obtained through various split sizes of top-20, top-10, and top-5 for the non-affective aware recommendations lists. The other five columns show the affective aware hit rate results. To make the regular non-affective aware recommender become an effective aware Recommender, choose the winner from the five-distance metric. In our case, the Cosine Similarity hit rate as shown in the "Cos" column yields the best result across the top-20, top-10, and top-5 when comparing among the others.

Table 10: Sampling on various split sizes for each test user to obtain the respective recommendations list and the reranking recommendations lists by five-distance metrics and hit rates on each user's category for reporting the average results.

| Top20 Split | Mid Hit% | Euc Hit% | Mht Hit% | Mki Hit% | Cos Hit% | Pear Hit% |
|---|---|---|---|---|---|---|
| 10-90 | 69.34 | 69.34 | 69.34 | 69.34 | 69.34 | 69.34 |
| 20-80 | 64.98 | 64.98 | 64.98 | 64.98 | 64.98 | 64.98 |
| 30-70 | 59.86 | 59.86 | 59.86 | 59.86 | 59.86 | 59.86 |
| 40-60 | 54.24 | 54.24 | 54.24 | 54.24 | 54.24 | 54.24 |
| 50-50 | 48.42 | 48.42 | 48.42 | 48.42 | 48.42 | 48.42 |
| 60-40 | 41.53 | 41.53 | 41.53 | 41.53 | 41.53 | 41.53 |
| 70-30 | 34.37 | 34.37 | 34.37 | 34.37 | 34.37 | 34.37 |
| 80-20 | 25.75 | 25.75 | 25.75 | 25.75 | 25.75 | 25.75 |
| 90-10 | 16.42 | 16.42 | 16.42 | 16.42 | 16.42 | 16.42 |
| Top5 Split | Mid Hit% | Euc Hit% | Mht Hit% | Mki Hit% | Cos Hit% | Pear Hit% |
| 10-90 | 82.26 | 69.41 | 69.41 | 69.67 | 69.80 | 69.80 |
| 20-80 | 78.20 | 64.30 | 64.13 | 64.00 | 65.87 | 65.57 |
| 30-70 | 72.46 | 59.57 | 59.67 | 59.51 | 60.23 | 60.23 |
| 40-60 | 67.11 | 54.10 | 55.02 | 54.33 | 54.33 | 54.03 |
| 50-50 | 60.98 | 49.84 | 49.51 | 49.38 | 48.10 | 47.87 |
| 60-40 | 53.34 | 42.13 | 42.03 | 41.87 | 40.82 | 40.79 |
| 70-30 | 44.03 | 34.46 | 34.69 | 34.59 | 34.69 | 34.13 |
| 80-20 | 34.56 | 25.48 | 25.54 | 25.28 | 25.41 | 26.07 |
| 90-10 | 21.80 | 17.11 | 16.67 | 17.18 | 15.41 | 16.00 |
| Top10 Split | Mid Hit% | Euc Hit% | Mht Hit% | Mki Hit% | Cos Hit% | Pear Hit% |
| 10-90 | 76.89 | 68.82 | 68.97 | 68.93 | 69.89 | 69.54 |
| 20-80 | 72.49 | 64.26 | 64.23 | 64.05 | 65.66 | 65.08 |
| 30-70 | 67.31 | 59.33 | 59.41 | 59.44 | 60.41 | 60.00 |
| 40-60 | 61.59 | 53.90 | 54.08 | 54.03 | 54.54 | 54.43 |
| 50-50 | 55.13 | 48.57 | 48.74 | 48.59 | 48.18 | 48.30 |
| 60-40 | 47.77 | 41.92 | 42.02 | 41.93 | 41.13 | 41.39 |
| 70-30 | 39.66 | 34.64 | 34.66 | 34.52 | 34.13 | 34.00 |
| 80-20 | 29.93 | 25.54 | 25.85 | 25.60 | 25.90 | 25.75 |
| 90-10 | 19.28 | 16.54 | 16.66 | 16.54 | 16.25 | 16.08 |

## 6 FUTURE WORK

We plan to elaborate our work in mvec and uvec from making the Recommender become emotion aware of the recommendations making process, a bottom-up approach, to build an Emotion Aware Recommender from the top down. We also plan to use affective features in users' emotion profiles to enhance Group Recommender in group formation, group dynamics, and group decision-making.

## 7 CONCLUSIONS

We illustrate a strategy to transform a non-affective aware Recommender to become affective aware. We started by developing an Emotion Detection and Recognition (EDR) model to classify seven emotional features in tweets through emotion tags stored in the Twitter database. We then transferred the EDR model's learning from classifying the affective features of tweets to classify a movie's emotion through the movie overview. We generated emotional features, mvec, for each collected movie from TMDb and joined it with the ratings dataset found in the MovieLens repository. We use the emotion labeled ratings dataset to make the top-N recommendations list through an SVD-CF Recommender while adding functions to support uvec and mvec for affective computing and analysis. We reranked all test users'users' top-N recommendations lists through five-distance metrics. We systematically evaluated the recommendations lists' performance and the reranked recommendation lists and found Cosine Similarity distance metrics performed the best. We conclude that the Cosine Similarity algorithm is the most suitable distance metrics to use in making a non-affective aware SVD-CF Recommender affective aware.

## REFERENCES

Bhowmick, P.K., Basu, A., Mitra, P., Prasad, A., 2009. Multi-label text classification approach for sentence level news emotion analysis, in: International Conference on Pattern Recognition and Machine Intelligence. Springer, pp. 261–266.

Canales, L., Martínez-Barco, P., 2014. Emotion detection from text: A survey, in: Proceedings of the Workshop on Natural Language Processing in the 5th Information Systems Research Working Days (JISIC). pp. 37–43.

De Silva, L.C., Miyasato, T., Nakatsu, R., 1997. Facial emotion recognition using multi-modal information, in: Proceedings of ICICS, 1997 International Conference on Information, Communications and Signal Processing. Theme: Trends in Information Systems Engineering and Wireless Multimedia Communications (Cat. IEEE, pp. 397–401.

Ekman, P., 2006. Darwin and facial expression: A century of research in review. Ishk.

Ekman, P., 1999. Basic emotions. Handbook of cognition and emotion 98, 16.

Ekman, P., Friesen, W.V., Ellsworth, P., 2013. Emotion in the human face: Guidelines for research and an integration of findings. Elsevier.

Fink, C.R., Chou, D.S., Kopecky, J.J., Llorens, A.J., 2011. Coarse- and Fine-Grained Sentiment Analysis of Social Media Text. Johns hopkins apl technical digest 30, 22–30.

Harper, F.M., Konstan, J.A., 2016. The movielens datasets: History and context. ACM Transactions on Interactive Intelligent Systems (TiiS) 5, 19. http://dx.doi.org/10.1145/2827872

Ho, I.L., Ai Thanh, Menezes, Tagmouti, Y., 2006. E-mrs: Emotion-based movie recommender system, in: Proceedings of IADIS E-Commerce Conference. USA: University of Washington Both-Ell. pp. 1–8.

IMDb, 2020. Top 100 Greatest Movies of All Time (The Ultimate List) [WWW Document]. URL http://www.imdb.com/list/ls055592025/ (accessed 5.17.20).

Kim, Y., 2014. Convolutional neural networks for sentence classification. arXiv preprint arXiv:1408.5882.

Konstan, D., Konstan, D., 2006. The emotions of the ancient Greeks: Studies in Aristotle and classical literature. University of Toronto Press.

Liu, T., 2020. Multi-class Emotion Classification for Short Texts [WWW Document]. URL https://tlkh.github.io/text-emotion-classification/ (accessed 5.18.20).

Maas, A.L., Daly, R.E., Pham, P.T., Huang, D., Ng, A.Y., Potts, C., 2011. Learning Word Vectors for Sentiment Analysis, in: Proceedings of the 49th Annual Meeting of the Association for Computational Linguistics: Human Language Technologies. Association for Computational Linguistics, Portland, Oregon, USA, pp. 142–150.

Makice, K., 2009. Twitter API: Up and running: Learn how to build applications with the Twitter API. O'Reilly Media, Inc.

Marres, N., Weltevrede, E., 2013. Scraping the social? Issues in live social research. Journal of cultural economy 6, 313–335.

Miller, G.A., 1995. WordNet: a lexical database for English. Communications of the ACM 38, 39–41.

Naqvi, N., Shiv, B., Bechara, A., 2006. The role of emotion in decision making: A cognitive neuroscience perspective. Current directions in psychological science 15, 260–264.

Pennington, J., Socher, R., Manning, C.D., 2014. Glove: Global vectors for word representation, in: Proceedings of the 2014 Conference on Empirical Methods in Natural Language Processing (EMNLP). pp. 1532–1543.

Plutchik, R., 2001. The nature of emotions: Human emotions have deep evolutionary roots, a fact that may explain their complexity and provide tools for clinical practice. American scientist 89, 344–350.

Poria, S., Gelbukh, A., Cambria, E., Yang, P., Hussain, A., Durrani, T., 2012. Merging SenticNet and WordNet-Affect emotion lists for sentiment analysis, in: 2012 IEEE 11th International Conference on Signal Processing. IEEE, pp. 1251–1255.

Sosa, P.M., 2017. Twitter Sentiment Analysis using Combined LSTM-CNN Models. Eprint Arxiv.

Strapparava, C., Valitutti, A., others, 2004. Wordnet affect: an affective extension of wordnet., in: Lrec. Citeseer, p. 40.

Tayib, S., Jamaludin, Z., 2016. An Algorithm to Define Emotions Based on Facial Gestures as Automated Input in Survey Instrument. Advanced Science Letters 22, 2889–2893.

TMDb, 2018. TMDb About [WWW Document]. URL https://www.themoviedb.org/about?language=en (accessed 5.11.18).

Tripathi, V., Joshi, A., Bhattacharyya, P., 2016. Emotion analysis from text: A survey. Center for Indian Language Technology Surveys.

Valitutti, A., Strapparava, C., Stock, O., 2004. Developing affective lexical resources. PsychNology Journal 2, 61–83.

Wilson, T., Wiebe, J., Hoffmann, P., 2005. Recognizing contextual polarity in phrase-level sentiment analysis, in: Proceedings of Human Language Technology Conference and Conference on Empirical Methods in Natural Language Processing. pp. 347–354.